# 4MOST Consortium Survey 1: The Milky Way Halo Low-Resolution Survey

Amina Helmi[1]
Mike Irwin[2]
Alis Deason[3]
Eduardo Balbinot[1]
Vasily Belokurov[2]
Joss Bland-Hawthorn[4]
Norbert Christlieb[5]
Maria-Rosa L. Cioni[6]
Sofia Feltzing[7]
Eva K. Grebel[8]
Georges Kordopatis[9]
Else Starkenburg[6]
Nicholas Walton[2]
C. Clare Worley[2]

[1] Kapteyn Instituut, Rijksuniversiteit Groningen, the Netherlands
[2] Institute of Astronomy, University of Cambridge, UK
[3] Department of Physics, Durham University, UK
[4] Sydney Institute for Astronomy, University of Sydney, Australia
[5] Zentrum für Astronomie der Universität Heidelberg/Landessternwarte, Germany
[6] Leibniz-Institut für Astrophysik Potsdam (AIP), Germany
[7] Lund Observatory, Lund University, Sweden
[8] Zentrum für Astronomie der Universität Heidelberg/Astronomisches Rechen-Institut, Germany
[9] Observatoire de la Côte d'Azur, Nice, France

The goal of this survey is to study the formation and evolution of the Milky Way halo to deduce its assembly history and the 3D distribution of mass in the Milky Way. The combination of multi-band photometry, Gaia proper motion and parallax data, and radial velocities and the metallicity and elemental abundances obtained from low-resolution spectra of halo giants with 4MOST, will yield an unprecedented characterisation of the Milky Way halo and its interface with the thick disc. The survey will produce a volume- and magnitude-limited complete sample of giant stars in the halo. It will cover at least 10 000 square degrees of high Galactic latitude, and measure line-of-sight velocities with a precision of 1–2 km s$^{-1}$ as well as metallicities to within 0.2 dex.

## Scientific context

Halo stars in the Milky Way spend significant amounts of time at large distances from the Galactic centre, hence their trajectories are sensitive to the mass distribution of the dark matter halo. Our survey, the 4MOST Consortium Milky Way Halo Low-Resolution Survey, is therefore key for measuring the full mass distribution of the Milky Way. In conjunction with the more local Milky Way Halo High-Resolution Survey, and the detailed surveys of the Milky Way disc and bulge carried out in the 4MOST MIlky Way Disc And BuLgE Low- and High-Resolution Surveys (4MIDABLE-LR and 4MIDABLE-HR), it also aims to determine the complete merger and assembly history of the Galaxy (see Christlieb et al., p. 26; Chiappini et al., p. 30 and Bensby et al., p. 35).

The Galactic halo contains large amounts of substructure at distances beyond 20 kpc, discovered with wide-field photometric surveys more than 10 years ago (Belokurov et al., 2006). At these large distances, debris is more spatially coherent because the mixing timescales are long. With the release of the Gaia astrometric data in April 2018 and the availability of 6D phase-space information, it has been demonstrated that the inner halo is dominated by merger debris from a single object as large as the Small Magellanic Cloud at the time of accretion (Helmi et al., 2018; Belokurov et al., 2018). Beyond our Galactic neighbourhood, the chemical characterisation of which is the focus of the Milky Way High-Resolution Survey (Christlieb et al, p.26), there is still much to learn. To make significant progress and pin down the full merger history of the Milky Way we require spectroscopic data of distant halo stars over a large portion of the sky that can be combined with the parallax and proper motion information from Gaia.

Most current mass estimates of the Milky Way have relied on small numbers of tracers, and hence are likely subject to bias given the substructures present in the halo. The existing data sets contain, at most, 150 objects beyond 50 kpc (dwarf galaxies, globular clusters, halo stars; see, for example, the review article by Bland-Hawthorn & Gerhard, 2016). Gaia will provide partial data for a few thousand stars out to 30–60 kpc, including tangential velocities with errors between 5 and 50 km s$^{-1}$ depending on their apparent magnitudes. 4MOST will push the radial velocities much further, down to $G \sim 20$ magnitudes at 1–2 km s$^{-1}$ accuracy. Our survey will also measure line-of-sight velocities for stars at the tip of the red giant branch up to a distance of 250 kpc, and the rarer carbon and other asymptotic giant branch stars to 1 Mpc. Not only will the estimates of the total mass of the Galaxy be much more precise, but also its 3D distribution (density and shape) will be within reach using various dynamical modelling techniques.

Our survey will also detect faint stellar streams and any substructures in a combined abundance-kinematic space. In particular, the follow-up of thin streams will allow us to pin down the lumpiness of the dark matter halo and to constrain its nature (Erkal & Belokurov, 2016; Bonaca et al., 2018). All of these measurements will thus lead to strong constraints on cosmological models.

A significant by-product of determining the metallicity distribution function across the halo will be the discovery of extremely metal-poor stars, which will be the focus of further, more detailed follow-up, most likely using facilities other than 4MOST.

## Specific scientific goals

The specific goals of our survey are:
– Determining the density profile, shape and characteristic parameters of the dark matter halo of the Milky Way — including testing alternative theories of gravity such as Modified Newtonian Dynamics (MOND) — and possibly their evolution in time.
– Measurement of the perturbations induced by clumps on the spatial and kinematic properties of cold streams leading to constraints on the mass spectrum of perturbers and on the nature of dark matter.
– Quantifying the amount of kinematic substructure as a function of distance and location on the sky. This will allow the discovery of substructures, new dwarf galaxies and other low surface brightness objects, the characterisation

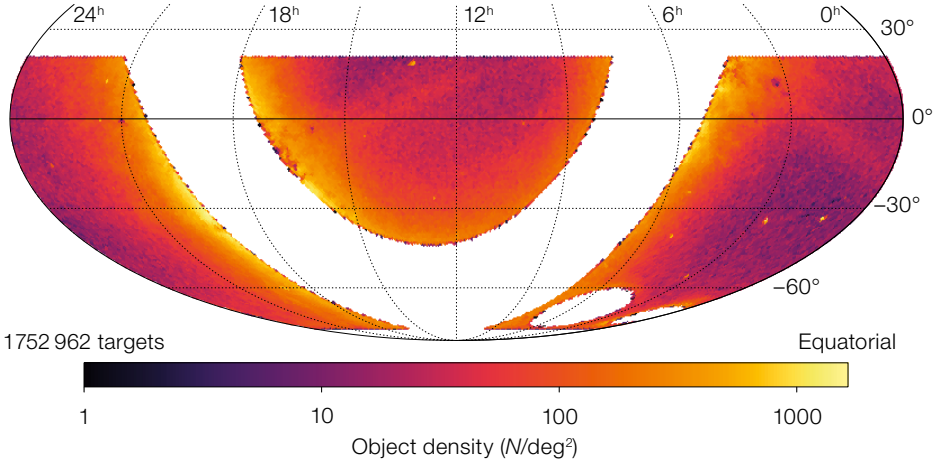

Figure 1. Input catalogue density distribution for the goal survey area of the Milky Way Halo Low-Resolution Survey. The stars shown here have been extracted from Gaia Data Release 2 (Gaia collaboration et al., 2018) and satisfy the following criteria: $-10 < G + 5\log_{10}(\text{proper motion}) < 10$, parallax $-2\sigma_{\text{parallax}} < 0.2$; $0.55 < G-G_{\text{RP}} < 0.8$ magnitudes, and $15 < G < 20$ magnitudes. This leads to a sample of approximately 2 million objects satisfying the declination range and $|b| > 20$ degrees, after pruning out 5- and 10-degree radius regions around the Small and Large Magellanic Clouds, respectively. Detailed coordination with the 4MOST Consortium Magellanic Cloud Survey (see Cioni et al., p. 54) will be carried out to ensure a smooth transition between the surveys, as well as refinement of the target selection criteria. Notice the Sagittarius streams and other halo over-densities in this version of the input catalogue for our survey. The goal survey area is $-80 < \text{dec} < +20$ degrees, however, it should be noted that the baseline survey for 4MOST is between $-70 < \text{dec} < +5$ degrees, and hence targets outside this footprint are less likely to be observed (see Guiglion et al., p. 17).

- of their properties and their relation to the build-up of the halo.
– Characterisation of the metallicity and elemental abundance distribution (mostly magnesium and iron) throughout the halo, and also of each of the individual structures discovered. This will yield enhanced samples of objects with very low metallicities or peculiar elemental abundances for more detailed follow-up, complementing the 4MOST Consortium Milky Way Halo High-Resolution Survey (Christlieb et al., p. 26) which focuses on the halo near the Sun. Such samples should constrain the properties and yields of the first generation of stars (Population III).
– Characterisation of the stellar halo-thick disc interface from overlap with the 4MIDABLE-LR survey (Chiappini et al., p. 30) with the aim of jointly constraining the temporal assembly and evolution of the thick disc and inner halo.

## Science requirements

The survey we propose will lead to a sample comprising on the order of 1.5 million giant stars in the halo (mainly K giant stars but also including the rarer A stars, particularly blue horizontal branch stars, together with M giant stars and carbon stars) across the virial volume of the Milky Way, with kinematics precise to 1–2 km s$^{-1}$ and overall metallicities precise to $\leq 0.2$ dex. This means observing all giant stars in the halo in the magnitude range $15 \leq G \leq 20$ magnitudes, including those in the lower density regions of halo dwarf galaxies and globular clusters. This magnitude range overlaps at the bright end with the Milky Way Halo High-Resolution Survey, which not only provides cross-checks on derived stellar properties, but also ensures full linkage between local and distant halo populations. Since the halo density is low, a wide-field instrument like 4MOST is essential to cover a large area in a reasonable time. The depth of our survey is perfectly suited to the goals, and matches exactly the depth reached with Gaia with useful proper motion information. The strongest constraints on the mass distribution come from streams covering large angles on the sky, again pushing for large area surveys. For stars with higher signal-to-noise (S/N > 25 per Å) useful constraints on the α-element abundances ([α/Fe] with error ≤ 0.1 dex), will be obtained and these are very important to further aid subdividing and characterising halo substructures (see for example, Hayes et al., 2018; Helmi et al., 2018).

The optimal streams for the determination of the Milky Way gravitational potential (mass, shape, time evolution and granularity) are thin and cold, and typically originate in objects with stellar mass smaller than a few times $10^5 M_\odot$. Models of galaxy formation (Cooper et al., 2010) predict about 50–100 such thin streams observable down to a $G$ magnitude ~ 20, across a 10 000 square degree region of the sky, and there may be many more from disrupted globular clusters (Bonaca et al., 2014; and see Malhan et al., 2018 for the first detections with Gaia DR2). A smaller area leads to a significant loss in the number of such streams and this impacts the determination of the mass distribution in the halo (Sanderson et al., 2015).

Radial velocity estimates will generally be obtained from a combination of the Mg b triplet (Mgb) and the near-infrared Ca II triplet (CaT) regions, which both contain sets of strong absorption lines, also easily detectable in stars with low metallicity. The velocity precision has to be 1–2 km s$^{-1}$ in order to measure the mean velocity in a field to approximately 500 m s$^{-1}$, which promises excellent constraints on the mass distribution in the Milky Way, and the dark matter granularity imprinted in the velocities of stream stars (for example, Bonaca et al., 2018). We note that accurate constraints on the Galactic potential can be obtained even if only limited proper motion information is available. If narrow streams are not distributed isotropically on the sky, for example as a consequence of infall along filaments, it will be important to complement the kinematic maps of streams with those from field stars.

## Target selection and survey area

The density of the stellar halo, and hence of the kinematic tracers, is low. Moreover, since the density profile of the halo drops rapidly, these tracers are rare at large distances. The expected average source density is 100–200 stars per square degree at a *G* magnitude ~ 20. A large survey area (minimum 10 000 square degrees; goal 150 000 square degrees) is also needed to find the rare, precious red giant branch stars near the virial radius of the Milky Way. Candidate halo giant stars will be selected on the basis of Gaia photometry, parallax and proper motion information possibly supplemented with photometry from ground-based imaging surveys (DES, SDSS, VST, PanSTARRS). Furthermore, we will also specifically target stars in streams known at the time of the survey, potentially down to the main-sequence turnoff to increase the number of objects and provide tighter constraints on the dynamics of the stream. Our aim is to target every star lying in a cold stream area in the available magnitude range, since we expect (field) contamination at a level of 70%–100%, depending on how the stars are pre-selected. Although Gaia will yield some useful prior constraints on the distances of stars in the inner halo, photometric distance estimates will need to be combined with the Gaia data both to improve the inner halo distances and to provide a distance proxy for the outer halo. Here in particular, the 4MOST spectra will be key. The spectra will be coupled with extant broadband photometry to derive photometric distances and accurate radial velocities for the halo star samples (for example, Xue et al., 2014). The same survey will also make full use of the combination of kinematic and positional information combined with chemical, i.e., [Fe/H] and [α/Fe], signatures to characterise substructures at large radii, be they streams or dwarf galaxies.

The goal survey area ranges from declinations (dec) of +20 to –80 degrees and covers all right ascensions (RAs) satisfying Galactic latitudes |*b*| > 20 degrees. This yields some 4500 square degrees north of the celestial equator and around 12 500 square degrees south of the equator, giving a total of 17 000 square degrees. As mentioned previously, galaxy formation simulations suggest a minimum requirement for the halo survey area to be at least 10 000 square degrees, while a desirable goal would be to cover at least 15 000 square degrees.

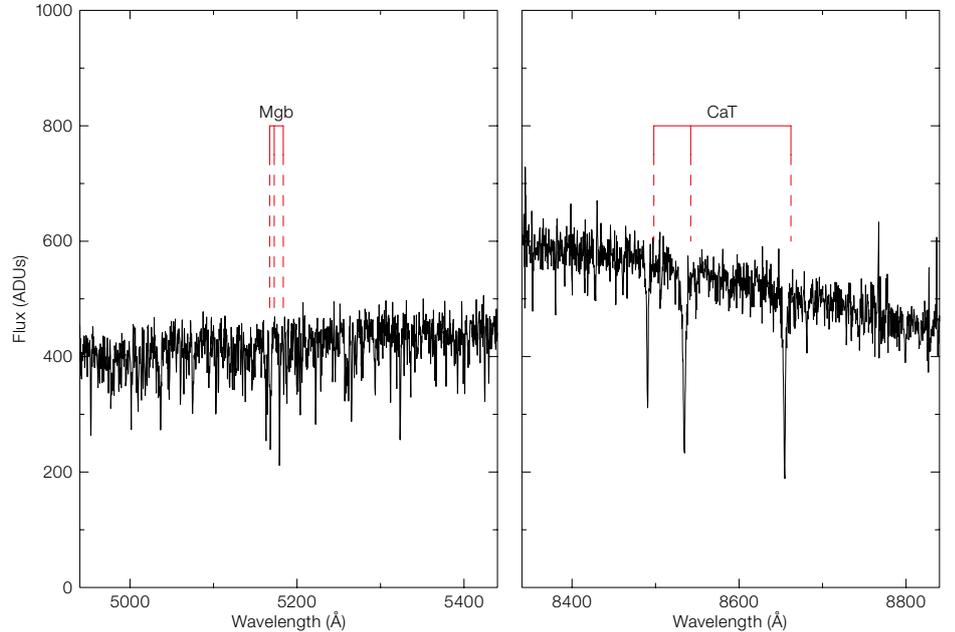

### Spectral success criterion and figure of merit

The spectral success criterion of our survey will not be binary (i.e., passed/failed), but "fuzzy" since spectra with S/N below the boundary value are still useful for deriving radial velocities, albeit with a lower precision. For computing the spectral success value, we will employ a non-linear function f(S/N) that maps the S/N of each spectrum onto the value range [0,1] and is defined to be 0.5 if the S/N = 10 per Å in the continuum in the Mgb and CaT regions (compare Figure 2).

The survey figure of merit (FoM) is chosen to yield a high completeness per field with a large fraction of the stars satisfying well-defined S/N constraints over specific wavelength regions. The overall survey FoM is defined to be

$$\mathrm{FoM} = \min\left\{1.0, \left[\frac{\Sigma_i A_i \min(CF_i/0.8, 1.0)}{15000.0}\right]\right\}$$

where $A_i$ is the area of 4MOST field *i* in square degrees and $CF_i$ the completeness fraction for a field, i.e., the fraction of stars satisfying the S/N constraints; 15 000 square degrees reflects our main survey area goal.

Figure 2. Example of a 1D-extracted spectrum of a halo K giant star (*g* = 18.73, *i* = 17.51 magnitudes) from a 4MOST simulated exposure of 3 × 1020 seconds in dark conditions. The stellar Mgb and CaT lines are blueshifted from their reference values by the high negative heliocentric velocity (–267 km s$^{-1}$) of the star. The average S/N in the continuum in these regions is around 25 Å$^{-1}$. According to the study by the Galactic pipeline working group, these can reach a precision of [Fe/H] ~ 0.15 and [α/Fe] ~ 0.1 dex.